\documentclass{aa}
\input epsf
\usepackage{graphics}
\usepackage{amssymb}
\usepackage{amsmath}
\def\Hmolec{{\hbox{\scriptsize H}_2}}
\def\kin{\hbox{\scriptsize kin}}
\def\molec{\hbox{\scriptsize mol}}
\def\innn{\hbox{\scriptsize in}}
\def\outtt{\hbox{\scriptsize out}}

\def\sec{\hbox{s}}

\def\cm{\hbox{cm}}
\def\km{\hbox{km}}
\def\kel{\hbox{K}}

\newcommand{\apj}{ApJ}

\newcommand{\aap}{A\&A}
\newcommand{\mnras}{MNRAS}

\newcommand{\be}{\begin{equation}}
\newcommand{\ee}{\end{equation}}
\newcommand{\onu}{_{\nu}}
\newcommand{\jbar}{\overline{J}}
\newcommand{\ba}{\begin{eqnarray}}
\newcommand{\ea}{\end{eqnarray}}

\def\email#1{{\tt #1}}
\newcommand{\comma}{\,,}
\newcommand{\fullstop}{\,.}
\begin{document}

\title{Numerical methods for non-LTE line radiative transfer:
Performance and convergence characteristics}
\titlerunning{Non-LTE line radiative transfer}
\authorrunning{van Zadelhoff et al.}

\author{G.-J. van Zadelhoff \inst{1} \and
 C.P. Dullemond \inst{2} \and
F.F.S. van der Tak \inst{3} \and
J.A. Yates\inst{4} \and
S.D. Doty  \inst{5} \and
V. Ossenkopf  \inst{6} \and
\\M.R. Hogerheijde  \inst{7} \and
M. Juvela \inst{8} \and
H. Wiesemeyer  \inst{9} \and
F.L. Sch\"{o}ier \inst{1}
} 

\offprints{G.J. van Zadelhoff, \email{zadelhof@strw.leidenuniv.nl}}
\institute{Leiden Observatory, P.O. Box 9513, NL--2300 RA Leiden, The
Netherlands \and Max Planck Institut f\"ur Astrophysik, Postfach 1317,
D--85741 Garching, Germany \and Max Planck Institut f\"{u}r
Radioastronomie, Auf dem H\"{u}gel 69, D--53121 Bonn, Germany \and
University College London, Gower Street, London WC1E 6BT, United
Kingdom \and Department of Physics and Astronomy, Denison University,
Granville, OH 43023, USA \and 1. Physikalisches Institut,
Universit\"{a}t zu K\"{o}ln, Z\"{u}lpicher Str. 77, D--50937,
K\"{o}ln, Germany \and Steward Observatory, The University of Arizona,
933 N. Cherry Ave.  Tucson AZ 85721, USA \and Helsinki University
Observatory, T\"{a}htitorninm\"{a}ki, P.O. Box 14, FIN-00014
University of Helsinki, Finland\and Institut de Radio-Astronomie
Millim\'{e}trique, 300 rue de la piscine, Domaine Universitaire,
F--38406 St.\ Martin d'H\`{e}res, France }

 \date{\today / Accepted ....}

\abstract{Comparison is made between a number of independent
computer programs for radiative transfer in molecular rotational
lines. The test models are spherically symmetric circumstellar
envelopes with a given density and temperature profile. The first two
test models have a simple power law density distribution, constant
temperature and a fictive 2-level molecule, while the other two test
models consist of an inside-out collapsing envelope observed in
rotational transitions of HCO$^{+}$. For the 2-level molecule test
problems all codes agree well to within 0.2\%, comparable to the
accuracy of the individual codes, for low optical depth and up to 2\%
for high optical depths ($\tau$=4800). The problem of the collapsing
cloud in HCO$^{+}$ has a larger spread in results, ranging up to 12\%
for the $J$=4 population. The spread is largest at the radius where
the transition from collisional to radiative excitation occurs. The
resulting line profiles for the HCO$^{+}$ $J$=4--3 transition agree to
within 10\%, i.e., within the calibration accuracy of most current
telescopes. {The comparison project and the results described in this
paper provide a benchmark for future code development, and give an
indication of the typical accuracy of present day calculations of
molecular line transfer.}}

\maketitle


\section{Introduction}
Molecular lines are excellent probes of the physical and chemical
conditions in interstellar clouds, protostellar envelopes,
circumstellar shells around late-type stars, photon-dominated regions
etc.  Furthermore, molecular line transitions play a key role in
probing the properties of galaxies and their evolution. The
interpretation of such lines requires the use of line radiative
transfer programs which can calculate accurately the {non-LTE (local
thermodynamic equilibrium)} level populations and the resulting output
spectra. See Black (2000) for a recent review.

It is known from stellar atmosphere research that subtle errors in
radiative transfer algorithms {can lead to significantly incorrect
results} (Mihalas 1978). A particularly well-known problem of this
kind is an insufficiently stringent convergence criterion at high
optical depths.  In the absence of a-posteriori checks on numerical
results, the best way to validate the methods is by the use of various
techniques, if possible with many independent codes of different
types.  Once the reliability and the behavior of the codes has been
established and understood, they can be safely used within the limits
of parameter space within which tests have been carried out.

In this paper, the setup and results of a large-scale campaign to
compare line radiative transfer programs are described. The test
problems are spherically symmetric, and can be modeled by a
1-dimensional radiative transfer code. Codes capable of handling more
than one dimension could be compared in a similar way in the future.

The main aim of this paper is to examine and compare the various
radiative transfer methods that are currently used in the
astrophysical community for modeling the shape and strength of
molecular lines emerging from interstellar and circumstellar
matter. {A series of problems is chosen that represent the
difficulties that may be encountered in comparison with data from
present-day and future ground-based telescopes, such as the
Sub-Millimeter Array (SMA)}, the Atacama Large Millimeter
Array (ALMA) and future airborne and space-borne {missions such}
as SOFIA and the Herschel Space Observatory. The test problems and
their solutions presented here are available to the community via a
web-site, allowing users to run the same problems, check the accuracy
of their codes and thus speed up the further development of their own
radiative transfer codes. It is hoped that this will stimulate a more
widespread use of these codes for the interpretation of molecular line
observations.

We present two test problems, each at two different optical
depths (i.e.~in total four problems). The first problem is a fictive
2-level molecule in a spherical envelope with a powerlaw density with
no systematic velocity and a constant temperature. This problem is
meant to tune all codes before examining the more complex, realistic
problem. The main test problem
is based on the 1-dimensional (1-D) inside-out collapse model, where
the level populations are computed for HCO$^{+}$ at various
abundances.  The HCO$^+$ ion is chosen as a representative example of
the molecule which samples gas with a large range in densities and is
readily observable in a variety of astrophysical regions.  Both a high
and a low optical depth model, representative of the main isotope
HCO$^+$ and the less abundant isotope H$^{13}$CO$^+$, are used to
check the convergence properties of each code. All test problems and
their results can be found at the webpage {\tt
http://www.strw.leidenuniv.nl/$\sim$radtrans}.

\section{Molecular line radiative transfer}
\label{sec: radtrans}

The radiative transfer problem is represented by an equation describing the
emission, absorption and movement of photons along a straight line in a
medium:
\begin{equation} 
\frac{d I_{\nu}}{ds}= j_{\nu} - \alpha_{\nu} I_{\nu} 
\fullstop
\label{eq: formrad} 
\end{equation} 
with the notation adopted from \cite{rybickilightman:1979}.  Equation
(\ref{eq: formrad}) is the differential description of the intensity
$I\onu$ along a photon path ($ds$) at frequency $\nu$, where $j_{\nu}$
[erg s$^{-1}$ cm$^{-3}$] and $\alpha_{\nu}$ [cm$^{-1}$] denote the
emission and absorption coefficients. Another way of writing
Eq. (\ref{eq: formrad}) is:
\begin{equation} 
\frac{d I_{\nu}}{d\tau_{\nu}}= S\onu-I\onu
\comma
\label{eq: form2rad}
\end{equation}   
where $S\onu={j\onu}/{\alpha\onu}$ is referred to as the source
function (the emissivity of the medium per unit optical depth), and
the optical depth $\tau_{\nu}$ is defined in differential form as
$d\tau_{\nu}= \alpha\onu\, ds$.  Equation (\ref{eq: form2rad}) can be
written in integral form, which is the form that is most often used in
radiative transfer codes:
\begin{equation}
I_\nu(\tau)= \int_{0}^{\tau} S\onu(\tau') e^{\tau'-\tau}d\tau'
\comma
\label{eq: form3rad}
\end{equation}
where $\tau$ is the optical depth between the point where $I_\nu$ is
evaluated and spatial infinity along the line (i.e.,~$s=-\infty$).
This integral is evaluated along all possible straight lines through
the medium. In practice, these will be a discrete sample of lines
covering space and direction as well as possible.

For the problem of molecular line transfer the emission and
absorption coefficients are determined by the transition rates between
the various rotational and/or vibrational levels of the molecule, and
the population of these levels. For a transition from level $i$ to
level $j$ (where the energy of level $i$ is greater than that of level
$j$) the emission and absorption coefficients are given by:
\begin{eqnarray}
j_{ij}(\nu) &=& n_i A_{ij} \phi_{ij}(\nu)\comma \\
\alpha_{ij}(\nu) &=& (n_j B_{ji} - n_i B_{ij}) \phi_{ij}(\nu)\comma
\label{eq: extinction}
\end{eqnarray}
where $n_i$ and $n_j$ [cm$^{-3}$] are the population densities of the
upper ($i$) and lower ($j$) level, and $A_{ij}$, $B_{ij}$ and $B_{ji}$
are the Einstein coefficients. The function $\phi_{ij}(\nu)$
represents the line profile for this transition, which is properly
described as a Voigt profile, a combination of a (micro-turbulent)
Gaussian and intrinsic Lorentzian line broadening.

The source function for a particular transition $S_{ij}$ is
independent of velocity if one assumes complete frequency
redistribution, i.e., the frequency deviation from line center of
absorbed and emitted photons are uncorrelated.  The source function
then becomes:
\begin{equation}
S_{ij}=\frac{j_{ij}(\nu)}{\alpha_{ij}(\nu)}=\frac{n_{i}
A_{ij}}{n_{j}B_{ji}-n_{i}B_{ij}}
\label{eq: sourcefunc}
\fullstop
\end{equation}

The relative level populations $n_{i}$ are determined from the statistical
equilibrium equation:
\begin{equation}
\begin{split}
\sum_{j>l}[n_{j}A_{jl}+(n_{j}B_{jl}-n_{l}B_{lj})\overline{J}_{jl}] \\
-\sum_{j<l}[n_{l}A_{lj}+(n_{l}B_{lj}-n_{j}B_{jl})\overline{J}_{lj}] \\
+\sum_{j}[n_{j}C_{jl}-n_{l}C_{lj}]=0
\comma
\end{split}
\label{eq: SE} 
\end{equation}
where $C_{ij}=n_{\rm col}K_{ij}$ with $K_{ij}$ the collisional rate
coefficients in cm$^{3}$ s$^{-1}$ and $n_{\rm col}$ the density of
collision partners, taken here to be H$_2$ in $J$=0.
$\overline{J}_{jl}$ is the integrated mean intensity over the line
profile:
\begin{equation}
\jbar_{ij}=\frac{1}{4 \pi}\int I\onu(\Omega) \, \phi(\nu) \, d\Omega \, d\nu
\fullstop
\label{eq: jbar}
\end{equation}
The symbol $\Omega$ represents the spatial direction in which the intensity
$I\onu(\Omega)$ is measured.

A useful concept for the analysis of radiative transfer
calculations is the excitation temperature
$T_{\mathrm{ex}}$ of the
transition between level $i$ and level $j$, given by
\begin{equation}
T_{\mathrm{ex}} =\frac{-h \nu_{ij}}{k} \left[\ln
\left(\frac{g_{j}}{g_{i}}\frac{n_{i}}{n_{j}}\right) \right]^{-1}
\comma
\label{eq: tex}
\end{equation}
where $k$ is Boltzmann's constant and $g_{i}$ the statistical weight
of level $i$. The energy difference between the two levels is given by
$h\nu_{ij}$. In local thermodynamic equilibrium (LTE),
$T_{\mathrm{ex}}$ equals the local gas temperature, while if
$T_{\mathrm{ex}}$ higher or lower, the excitation is super- or
sub-thermal. In addition, the intensity is proportional to
$T_{\mathrm{ex}}$ in the optically thin limit.
\begin{figure}
\resizebox{\hsize}{!}{\includegraphics{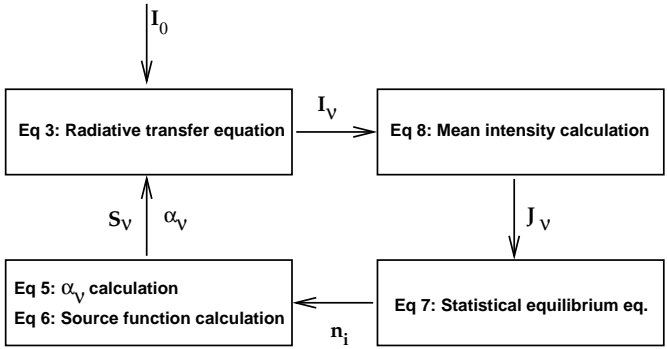}}
\caption{Flow diagram of the molecular line radiative transfer
problem. Since the problem is coupled (follow arrows), several iterations are
needed to calculate the true level populations $n_{i}$. All adopted symbols
are explained in \S \ref{sec: radtrans} }
\label{fig: radtrans}
\end{figure}

Equations (\ref{eq: form3rad})-(\ref{eq: jbar}) form a coupled set of
equations. The way in which the quantities depend on each other is
depicted in Fig.~\ref{fig: radtrans}. The intensities are found by
integrating the source function (Eq. 3). The source function and
extinction coefficients depend on the level populations
(Eqs. 4--5). These in turn depend on the intensities (Eqs. 7--8).  To
solve this set of equations one must determine the radiation field and
the level populations simultaneously. Since the radiation field
couples the level populations at different spatial positions to each
other through the transfer integral (Eq.~\ref{eq: form3rad}), the only
way of solving this system directly seems to be through complete
linearization and solving a huge matrix equation involving all level
populations at all spatial positions. In practice, the evaluation of
the matrix elements and the inversion of the problem consumes far too
much CPU-time as well as memory and is therefore beyond current
computing capabilities. An alternative and much simpler way is to
iteratively evaluate all equations, following the arrows, as
illustrated in Fig.~\ref{fig: radtrans}. This method is called
``Lambda Iteration'' and includes formal integral codes as well as
Monte Carlo methods.  Though simple, its principal disadvantage is
that it converges slowly at high optical depths. Many radiative
transfer codes therefore use a hybrid scheme: the direct inversion of
a simplified subset of the equations, and iteration to solve the
remaining problem. In this paper we have used codes of the Lambda
Iteration type (including Monte Carlo methods), of the ``Complete
Linearization'' type, and hybrid schemes such as
``Approximate/Accelerated Lambda Iteration'' and ``Accelerated Monte
Carlo''.

\section{Methods and codes}
\label{sec: desc}
In this paper {four} methods and eight different codes are
compared. Any practical method discretizes the problem to determine
the level populations at each position. A short description is given
in this section of each of these codes. {The first method (LI) is
not used, but given as an introduction for the other methods.} In
Table \ref{tab: codes}, the different codes and principal authors are
indicated. Detailed descriptions of each of the codes can be found in
the references given.
\subsection{Lambda Iteration (LI)}
The Lambda Iteration method is the most basic method of all. It involves the
iterative evaluation of level populations and intensities until the system
has converged. The name ``Lambda Iteration'' originates from the fact that
the process of iteration can be mathematically written into a formalism
involving a ``Lambda Operator''. This Lambda Operator represents the entire
procedure of computing the line-weighted mean intensities $\jbar_{ij}$ from
the source function. It involves the formal integral Eq.~(\ref{eq: form3rad})
along all possible lines through the medium, and includes the
angle-frequency integrations of Eq.~(\ref{eq: jbar}) to obtain the
$\jbar$. The Lambda Operator is defined as:
\begin{equation}
\jbar_{ij} = \Lambda_{ij} [S_{ij}]
\comma
\end{equation}
and is therefore a global operator. The $\Lambda_{\nu}$ operator is a
matrix operator connecting all points and all levels to each other.
To solve the radiative transfer numerically, the problem has to be
discretized both in space and frequency and initial level populations
have to be assumed. With these level populations, the Lambda Operator
is constructed by solving the radiative transfer for $I_{\nu}$ and
$\jbar$, which is inserted in Eq. (\ref{eq: SE}) to calculate the new
level populations. The procedure is repeated until the relative change
in the level populations or mean intensities between two successive
steps falls below a desired convergence criterium. The time to
reach convergence of
the solution is proportional to $\tau^2$ for $\tau\gg 1$ and is
therefore extremely slow for highly optically thick regions.  The very
small local changes in the level populations in successive iterations
could then be easily mistaken for convergence.

\subsection{Monte Carlo (MC)}
The Monte Carlo (MC) method is based on the simulation of basic
physical processes with the aid of random numbers. This makes the
formalisms of these codes relatively simple and intuitive as one has
to deal only with the basic formulae. But one must take proper care of
the statistics and be sure that all regions in space and frequency are
well sampled by the randomly distributed photon packages.  The Monte
Carlo method for molecular line transfer has been described by
\cite{bernes} for a spherically symmetric cloud with a uniform density.
A major advantage of MC codes is the possibility of non-regular
grids, in particular for multi-dimensional problems. In fact, MC
methods are straightforward to extend from 1-D to 2-D (Hogerheijde \&
van der Tak~2000) and even 3-D (Park \& Hong 1998, Juvela 1997). One
of the major disadvantages of the method is the long CPU time which is
needed to lower the random error intrinsic in the method. This
error decreases inversely proportional to the square root of the
number of simulated photons. In addition, the method suffers
from similar convergence problems as Lambda Iteration, as it is
formally a variant of the Lambda Iteration method.

Thus far, the Monte Carlo method has been implemented in two
ways.  In one set of codes, the radiation field is randomly sampled by
discrete photons that are emitted and absorbed in the material. In the
other set, the radiation field is sampled by random directions along
which the radiative transfer is evaluated.

\subsection{Approximated/Accelerated Lambda Iteration (ALI)}
This scheme is similar to the Lambda iteration scheme, with the
difference that the equations are pre-conditioned to speed up the
convergence. The simplest (but still highly efficient) way to
implement the ALI scheme is by splitting the Lambda Operator into the
local self-coupling contribution and the remainder:
\begin{equation}
\Lambda_{ij}=(\Lambda_{ij}-\Lambda_{ij}^{*})+\Lambda_{ij}^{*}
\end{equation}
with $\Lambda_{ij}$ the lambda operator between the $i$ and $j$ levels
and $\Lambda_{ij}^{*}$ the diagonal (local), or tridiagonal
(local+nearest-neighbor), part of the full Lambda operator. In
general more reliable convergence properties are obtained with
tridiagonal or higher bandwidths (Hauschildt et al. 1994). Since an
approximate operator $\Lambda_{ij}^{*}$ of this kind is easily
invertible, and since its matrix elements are relatively easy to
calculate, one can use this operator for the matrix inversion, and use
the iteration for solving the remainder (non-local part) of the
transfer problem. Since the self-coupling of a cell at high optical
depth is the bottle-neck that slows down convergence in Lambda
Iteration codes, this removal of the local self-coupling by direct
inversion of $\Lambda_{ij}^{*}$ can speed up the convergence
drastically. A full description of the method was given by, e.g.,
Rybicki \& Hummer (1991, 1992).

In addition to this operator splitting technique, one can apply
certain iteration-improvement schemes such as the Ng-acceleration
method (\cite{ngacc}). These are methods that can improve the
convergence of any linearly converging iteration scheme. In the
Ng-method one uses the previous results to estimate the convergence
behavior of the problem. After every four iteration steps an
extrapolation can then be performed towards the expected
convergence. The number of iterations is not a strict requirement but
four is typically found to give a reliable and significant
acceleration. This scheme is very effective as can be seen in Figs. 1
to 3 in Rybicki \& Hummer (1991), where the number of iterations is
plotted versus the convergence.

\subsection{Accelerated Monte Carlo (AMC)}
The difference in Accelerated Monte Carlo and Monte Carlo 
methods can be similarly described as the ALI compared to LI
method. Most important is the splitting of the mean intensity in an
external field and a local contribution. Formally,
\begin{eqnarray}
\jbar&=&(\Lambda-\Lambda^{*})[S^{\dagger}_{ul}(\jbar)]+\Lambda^{*}[S_{ul}(\jbar)] 
\\
&=&\jbar^{external}+\jbar^{local} \\
&=&\frac{1}{N} \sum_{i} I_{0,i} e^{-\tau_{i}}+\frac{1}{N} \sum_{i} 
S_{ul}[1-e^{-\tau}]
\end{eqnarray}
with S$^{\dagger}_{ul}(\jbar)$ the results from the previous iteration
and $N$ the number of photons.  This is an important issue as in
the standard Monte Carlo approach, most time is taken by photons
trapped in an optically thick cell, where due to absorption and
randomly directed emission the photons follow a random walk path
through the cell, with one step for each iteration.

The splitting can be done in different ways. Juvela (1997)
implements a diagonal lambda operator in Eqs. (12)-(14) by counting
explicitly those photons which were emitted and absorbed within the
same cell. A reference field described by Bernes et al. (1979) is used
to decrease the random fluctuations caused by the Monte Carlo
sampling.  Hogerheijde \& van der Tak (2000) perform sub-iterations to
calculate the local contribution in each cell, thus ensuring that in
each cell the local field and populations are always consistent with
the external field due to all other cells.  A third possibility of
acceleration, adopted by Sch\"{o}ier (2000), is the use of
core-saturation (Rybicki 1972, Hartstein \& Liseau 1998), where
optically thick line center photons are replaced by the local source
function.

\subsection{Local Linearization (MULTI type codes)}
This method was developed by \cite{sc85} and \cite{h94} to produce the
MULTI and SMULTI line radiation transport codes respectively.  This
approach perturbs Eqs. (2) to (8) linearly, neglecting second order or
higher terms.  The linearization greatly reduces the effect of high
optical depth terms and allows rapid convergence.  The MULTI and
SMULTI codes use the Olson diagonal approximate operator
(\cite{olson}) scheme to save on storage.  This operator uses the
diagonal of the $\Lambda$ matrix when computing the solution to the
linearized equations.  Although this adds in approximations and delays
convergence, it greatly reduces the required storage capacity and so
for many problems it allows the problem to be tackled in the first
place.  Olson's scheme simply assumes that the changes in the
intensity are related to the changes in the source and opacity terms
(the level populations).  The main difference to the other schemes is
that the solution to the linearized equations returns changes, $\delta
n$ and $\delta J$ to the level populations and the mean intensities
respectively.  There is always the possibility that a converged
solution contains unphysical negative populations, and this provides a
useful indicator for poor sampling and errors. In the version used
here a ALI-type scheme to complement the MULTI is used, providing
numerical stability but also slower convergence in optically thick
media.

\subsection{Radiative transfer codes}
\begin{table*}[ht!]
\begin{center}
\caption{Codes used}
\begin{tabular}{lllll}
\hline
Label$^{1}$ & Author & method & dimensions  & Reference \\
\hline
A &Juvela      & AMC & 1D, 2D \& 3D &  \cite{1997A&A...322..943J}\\  
B &Hogerheijde \&& AMC & 1D \& 2D &  Hogerheijde \& van der Tak (2000)  \\
&van der Tak & \\
C&Sch\"{o}ier & MC  & 1D  & Sch\"{o}ier (2000) \\
D&Doty        & ALI & 1D      &   \\
E&Ossenkopf   & ALI & 1D \& 2D      & Ossenkopf, Trojan \& Stutzki (2001) \\
F&Dullemond   & ALI & 1D \& 2D & Dullemond \& Turolla (2000)  \\
G&Yates       &(S)MULTI& 1D        & Rawlings \& Yates (2001) \\
H&Wiesemeyer & ALI & 1D \& 2D   & Wiesemeyer (1997)\\
\hline
\label{tab: codes}
 \end{tabular} \\[-0.3cm]
$^{1}$ Label used in Figures 4 and 6 for each of the codes.
\end{center}
\end{table*}

\medskip

\medskip
Each code described in this paper uses one or more of the above
techniques to calculate a converged set of level populations.  In this
section we list the codes which use each of the convergence methods
listed above.  Then for each code we describe how convergence is
accelerated, how each code samples the volume under consideration, and
state the convergence criteria used by each code.
\medskip

\begin{enumerate}
\item{Monte Carlo (MC):
\begin{itemize}
\item{ F. Sch\"{o}ier (\cite{2000PhDT.........6S}); The rate of
convergence depends on the number of model photons and the iterative
procedure. In the problems presented here, the counters of stimulated
emission are reset after each set of 5 iterations.  The number of
iterations needed for convergence in the ``classical" Monte Carlo
scheme is of the same order as the maximum optical depth in the model.
The core saturation method is included to speed up the convergence in
the high optical depth case.  To reach an accuracy of $\sim$10$^{-2}$
in the derived level populations, $\sim10^{5}-10^{6}$ model photons
per iteration are generally needed in the optically thick case.}
\end{itemize}}
\item{Accelerated Lambda Iteration (ALI): 
\begin{itemize} 
\item{V. Ossenkopf (\cite{2001A&A...378..608O}); All discretizations
are done by the code such that sufficiently small differences between
the grid points are guaranteed. This concerns the radial shells, the
grid of rays, the frequency grid, and the number of levels used. The
convergence is measured in terms of radiative energy densities, not
level populations, which are extrapolated from the Auer acceleration
scheme. Locally, the code may take into account turbulent sub-clumping
using the statistical description of \cite{1984MNRAS.208...35M}, but
this is not used here.
}
\item{S. Doty; the input data are interpolated onto the spatial grid
    of the problem with a variable method (log/linear/combined).
    This, coupled with extreme care in the ray-tracing integration,
    helps to ensure high accuracy. A local approximate lambda operator
    and Ng acceleration are used to solve both the line and continuum
    transfer. Convergence is determined by changes in both
    level populations and the local net heating rate, with small ($
    \le 10^{-4}$) changes required for the entire spatial grid for
    both accelerated and non-accelerated iterations.  }
\item{C.P.~Dullemond (\cite{2000A&A...360.1187D}); The formal
integration proceeds via a short-characteristics method using the
three-point quadrature formula of \cite{olsonkunasz:1987}. The
discrete local angles $\mu_i$ are chosen using the tangent-ray
prescription. This means that successive short characteristics match
up and no interpolations are required. At every grid point there are
41 angular points in $\mu$ with three extra around $\mu\simeq 0$ to
prevent undersampling.  Using the ALI method with a local operator and
Ng acceleration, the system is iterated towards a solution with a
convergence criterion $10^{-6}$ in level population.}
\item{H.\ Wiesemeyer (\cite{1997PhDT........21W}; The rate
of convergence is improved by applying the method of minimization of
residuals (Auer 1987).  Other acceleration methods, such as vector and
matrix extrapolation, are currently being implemented.  The solution
to the equation of radiative transfer is evaluated by various
numerical methods (either quadrature rules, or multi-value methods),
according to the accuracy required by the problem to be solved. The
code uses long characteristics, to preserve an isotropic distribution
of rays at any point of the model volume, following the multivariate
quadrature rule of Steinacker et al. (1996). Its performance is thus
rather optimized to solve smooth problems in more than one
dimension. All discretizations are performed by the code.}
\end{itemize}}
\item{Accelerated Monte Carlo (AMC): 
\begin{itemize}
\item{M. Juvela (Juvela 1997); This code uses the reference field
method to reduce random fluctuations. The velocity is discretized into
50 channels and the number of rays per iteration was taken to be
350. The ray generation is weighted such that the same number of rays
was shot through each annulus. The random number generators are reset
after each iteration, making it possible to use Ng acceleration on
every third iteration starting with the fifth iteration.  Convergence
was tested only for the eight lowest levels down to 1$\times10^{-3}$
in level population.}

\item{M. Hogerheijde \& F. van der Tak (\cite{2000A&A...362..697H});
{In this code, one cell at a time is considered, with $N$ rays
entering the cell from random directions. The radiative transfer is
followed along each of these rays, starting with the CMB field at the
edge of the cloud. In this way it is possible to calculate separately
the local contribution to the radiation field in the target cell,
allowing significant reduction of the computing time for optically
thick cells. For a first order estimate of the radiation field, the
same set of random numbers is used to describe the ray directions for
each iteration, thus resembling a fixed-ray (ALI) code. This process
yields a solution free of random noise but with possible inadequate
sampling of directions and velocities. In the second stage a different
set of random numbers is used for each iteration, providing true
random sampling of the radiation field. The number of rays is
increased for each cell until convergence is reached, ensuring proper
angular sampling of the radiation field everywhere. The equations of
statistical equilibrium are solved in each iteration to a fractional
error of $10^{-6}$ in all levels except the highest. The user
specifies a signal-to-noise ratio $S$; the MC noise is reduced by
increasing the number of photons such that the fractional error of the
levels between iterations is smaller than 1/$S$. For the test problem,
$S$=100 was used.}}

\end{itemize}}
\item{Complete Linearization (MULTI-type):
\begin{itemize}
\item{J. Yates (\cite{2001MNRAS.326.1423R}); In the problems
presented, the SMMOL code used 400 rays through the cloud to compute
the intensities at each radial grid-point in the spherical
cloud using a finite difference method to compute the intensities. The
grid sampling along each ray can adapt simply to take account of
large changes that affect the optical depth, typically caused 
by velocity changes along the ray.  In this
calculation, typically 100 grid-points were used to sample the full
absorption profile.  The SMMOL codes uses a convergence criterion of
$1\times 10^{-4}$ for both the level populations and the mean
intensities.  }
\end{itemize}
}
\end{enumerate}

\section{Description of the test models}

\subsection{Models 1a/1b: a simple 2-level molecule}
The first two test problems (problems 1a/1b) have a simple and
cleanly defined setup without velocity gradients and using a constant
temperature and line width.  Complicating factors such as a different
treatment of the spatial grid and frequency sampling are thus kept to
a minimum, so that every code should in principle be able to solve
these problems down to their specified accuracy.

The simple 2-level molecule test setup consists of a spherically
symmetric cloud with a power law hydrogen number density specified by
\begin{equation}
n_{\Hmolec}(r) = n_{\Hmolec}(r_0)\;\left( \frac{r}{r_0} \right)^{\alpha}
\quad \cm^{-3}
\comma
\end{equation}
where $r$ is the radius in $\cm$. We take $n_{\Hmolec}(r_0)=2.0\times
10^{7}\,\cm^{-3}$, $r_0=1.0\times 10^{15}\,\cm$ and $\alpha=-2$. The
kinetic temperature of this problem is taken to be constant:
$T_{\kin}(r) = 20\;\kel$. The abundance of the molecule (which we will
specify below) is constant as well: $X_{\molec}(r) \equiv
n_{\molec}(r)/n_{\Hmolec}(r) = X_{\molec} $, and the systematic
velocity is zero everywhere. The spherically symmetric cloud ranges
from $r_{\innn}=1.0\times 10^{15}\,\cm$ to $r_{\outtt}=7.8\times
10^{18}\,\cm$. For $r<r_{\innn}$ and for $r>r_{\outtt}$ the density is
assumed to be zero. The incoming radiation at the outer boundary is
the $T=2.728\;\kel$ microwave background radiation.

We choose a fictive 2-level molecule which is specified by
\begin{xalignat}{1}
E_2-E_1 &= 6.0\; \cm^{-1} = 8.63244 \;\kel \\
g_2/g_1 &= 3.0 \\
A_{21}  &= 1.0\times 10^{-4}\; \sec^{-1} \\
K_{21}  &= 2.0\times 10^{-10}\; \cm^3\,\sec^{-1}
\end{xalignat}
in which the downward collision rate is
$C_{21}=n_{\Hmolec}K_{21}\;\sec^{-1}$.  The total (thermal+turbulent) line
width $a$ is given by $a = 0.150\; \km\,\sec^{-1}$. The line profile is
assumed to be a Gaussian:
\begin{equation}\label{eq-turb-lineprof-def}
\phi(\nu) = \frac{c}{a\nu_{0}\sqrt{\pi}} 
\exp\left(-\frac{c^2(\nu-\nu_{0})^2}{a^2\nu_{0}^2}\right)
\end{equation}
where $c$ is the speed of light in units of $\km\;\sec^{-1}$, and 
$\nu_0$ is the frequency at line center.

We solve the problem for two different abundances:
\begin{alignat}{2}
\hbox{Problem 1a:} & \qquad & X_{\molec} &= 1.0\times 10^{-8}\nonumber\\
\hbox{Problem 1b:} & \qquad& X_{\molec} &= 1.0\times 10^{-6}\nonumber
\end{alignat}
Problem 1a is a simple case with a moderate optical depth ($\tau\simeq
60$ from star to infinity at line center). Problem 1b has much higher
optical depth ($\tau\simeq 4800$) and is therefore numerically
stiffer.

This problem was originally described and worked out by Dullemond
\& Ossenkopf (see Dullemond 1999).

\subsection{Models 2a/2b: A collapsing cloud in HCO$^{+}$}
The second two problems (problems 2a/2b) are examples of problems
typically encountered in the field of sub-millimeter molecular line
modeling. These problems have much of the complicated physics
included, such as velocity- and temperature gradients, non-constant
line widths, multiple levels etc. Moreover, the problem is based on a
numerical model provided externally, and therefore introduces the
complication of a given intermediate-resolution spatial grid. The
resulting spread of the results therefore gives a good indication of
how accurate the predicted line strengths and shapes for a typical
every-day-life model calculation are.

The model on which these ``realistic'' test problems are based was
constructed by Rawlings et al.~(1992, 1999) to analyze HCO$^+$ data
for an infalling envelope around a protostar. The prototypical example
of such a `class 0' young stellar object is B335. The model describes
how the cloud core collapses from the inside-out. Starting from a
nearly isothermal sphere in pressure balance, a perturbation triggers
the center of the cloud to collapse. This sends out a rarefaction wave
at the local speed of sound, which causes the outer parts of the cloud
to collapse as well. The model is similar to the analytical inside-out
collapse model by \cite{1977ApJ...214..488S}, but includes more
realistic physics. No rotation is assumed, which makes it ideal for a
1-D spherical comparison.

\begin{figure}
\resizebox{\hsize}{!}{\includegraphics{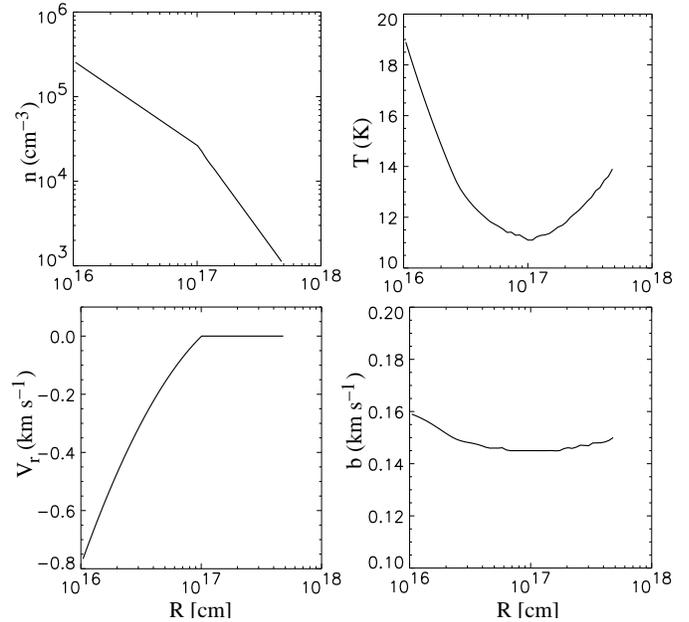}}
\caption{Physical structure of the test model 2. Upper left: density,
lower left: velocity, upper right: temperature, lower right: turbulent
line width $b$ [km s$^{-1}$]. Changes in the density and velocity
distributions are seen at the point where infall starts. The
temperature of the gas at the inside rises due to the infall while the
outside is heated by the interstellar radiation. }
\label{fig: model}
\end{figure}

Figure ~\ref{fig: model} shows the structure of the cloud at a
particular time during the collapse phase. This is the input model for
our test cases. The collapse can clearly be seen in the radial
velocity, which is 0 for radii greater than 10$^{17}$~cm, and directed
towards the source for smaller radii. The density profile is given by
a power-law of the form $n(r)=n_{0} (r/r_{0})^{m}$, where $m=-1.5$
inside the collapsing sphere ($r<$ 10$^{17}$ cm) and $m=-2.0$ outside.
The model parameters are specified at 50 radii logarithmically spaced
between $10^{16}$ and $4.6\times 10^{17}$ cm.

\noindent

The infall of a cloud has a significant influence on the molecular
line emission. The lines will be skewed to the blue, or double peaked
with a stronger blue-shifted component, depending on the velocity
field and optical depth of the line (e.g. Myers et
al. 2000). This feature can be intuitively understood because the
foreground gas has a lower excitation temperature than the background
gas. This only holds when the foreground gas is optically thick enough
in the line.  Otherwise a single-peaked lineshape, centered on the
source velocity, results.

 For the comparison presented here, the ion HCO$^{+}$ is used, where
an abundance compared to H$_{2}$ is adopted of 1$\times$10$^{-9}$ for
problem 2a, and 1$\times$10$^{-8}$ for problem 2b. This causes the optical
depths $\tau$ for the lowest 4 transitions to range from 0.1 to 10 for
problem 2a, and from 1 to 100 for problem 2b.  A common file of Einstein-$A$
coefficients and collisional rate coefficients of HCO$^+$ with H$_2$
in $J$=0 (\cite{1985MNRAS.214..419M}, \cite{1975ApJ...201..366G}) has
been used in all models. These can be downloaded from {\tt
http://www.strw.leidenuniv.nl/$\sim$radtrans}.  21 levels of the
HCO$^+$ ion are included in the model. Since the rate coefficients are
temperature dependent, all downward (upper$\rightarrow$lower) rate
coefficients are given for a number of temperatures (10, 20, 30,
40~K). At each temperature, a linear interpolation is performed to
calculate the local downward collisional rate coefficients.
Performing this interpolation accurately is extremely important
because the exponential relation between the excitation and
de-excitation coefficients can introduce large deviations. The upward
(lower$\rightarrow$upper) collisional rate coefficient is subsequently
calculated using the detailed balance relation,
\begin{equation}
\frac{C_{lu}}{C_{ul}}=\frac{g_{l}}{g_{u}} e^{E_{ul}/kT}.
\end{equation} 

The fact that the molecular data, in particular the collisional rate
coefficients, are not known with infinite precision introduces an
additional uncertainty in the solution of the radiative transfer
equations. The use of different collisional rate coefficients has an
impact on the level populations and thereby on the strength of the
predicted emission lines.  However, the spread in available rate
coefficients in the literature has no effect on the comparison
presented here, as all the codes use the same molecular data as
input. On a webpage {\tt http://www.strw.leidenuniv.nl/$\sim$moldata}
molecular files for a large number of molecular species will be made
available in the future (Sch\"oier et al., in preparation), including
a comparison of different rates from the literature and their effect
on the model results.

With these sets of problems the different codes are tested on the
following features:
\begin{itemize}
\item{Convergence of the code}
\item{Radiative transfer in optically thin lines}
\item{Radiative transfer in optically thick lines}
\item{Sampling of the radiation transfer in the presence of a velocity
field}
\item{Incorporation of the cosmic microwave background radiation
($T_{\rm CMB}$=2.728 K), which influences the lower levels near the
outside of the cloud.}
\end{itemize}

\section{Results}

\subsection{Results of problems 1a/1b}

Test problems 1a and 1b are relatively simple in the sense that no
velocity gradients and temperature variations are present. Also, since
the problem setup is specified analytically, the grid resolution can be
chosen to be as high as required for the particular code the
participant is using. By virtue of the zero systematic velocity, the
treatment of the inner boundary is not much of an issue, because the
line is perfectly thermalized at that radius. These test problems
therefore purely test the radiative transfer as such.

\begin{figure}
\begin{center}
\resizebox{7cm}{!}{\includegraphics{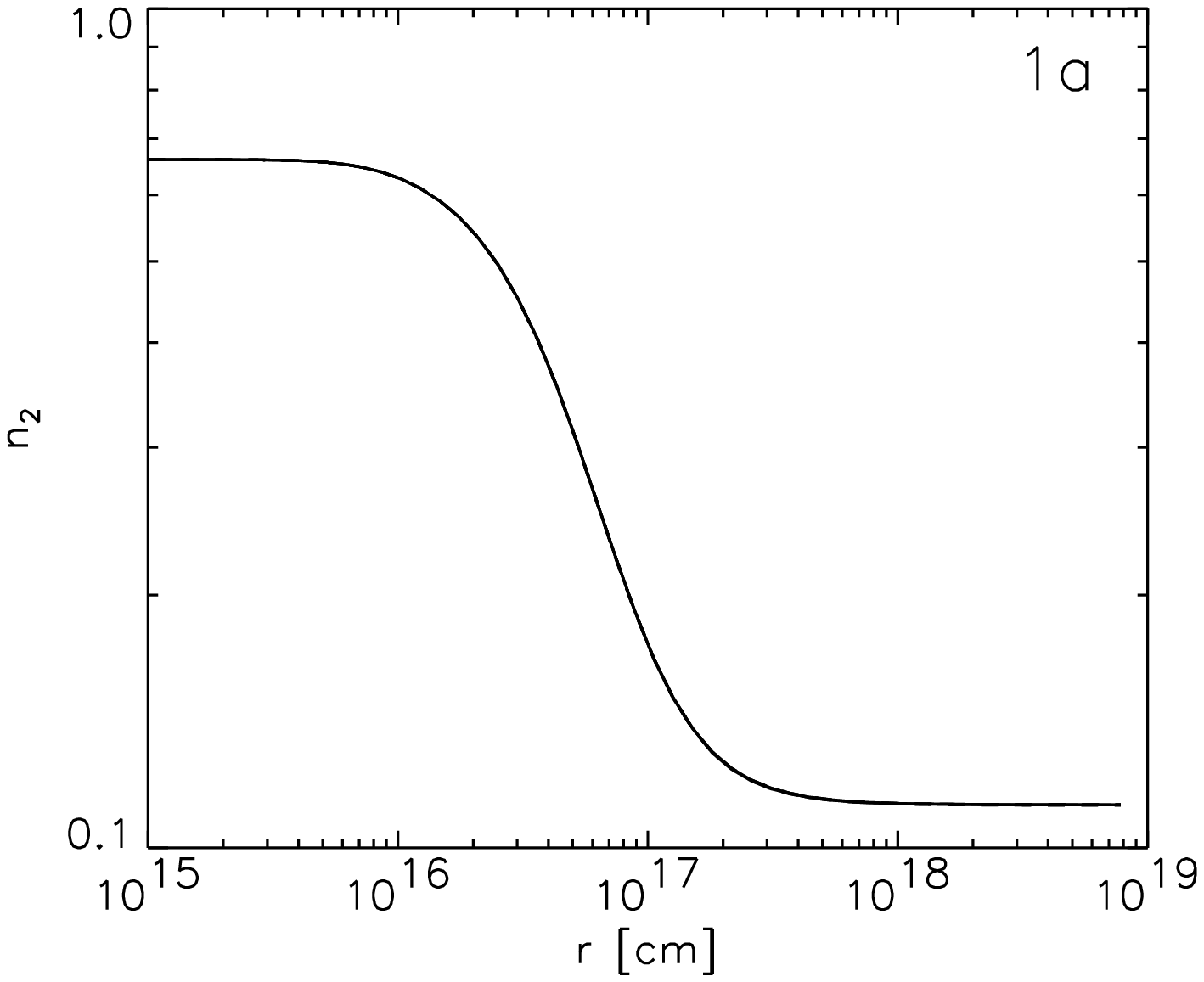}} \\
\resizebox{7cm}{!}{\includegraphics{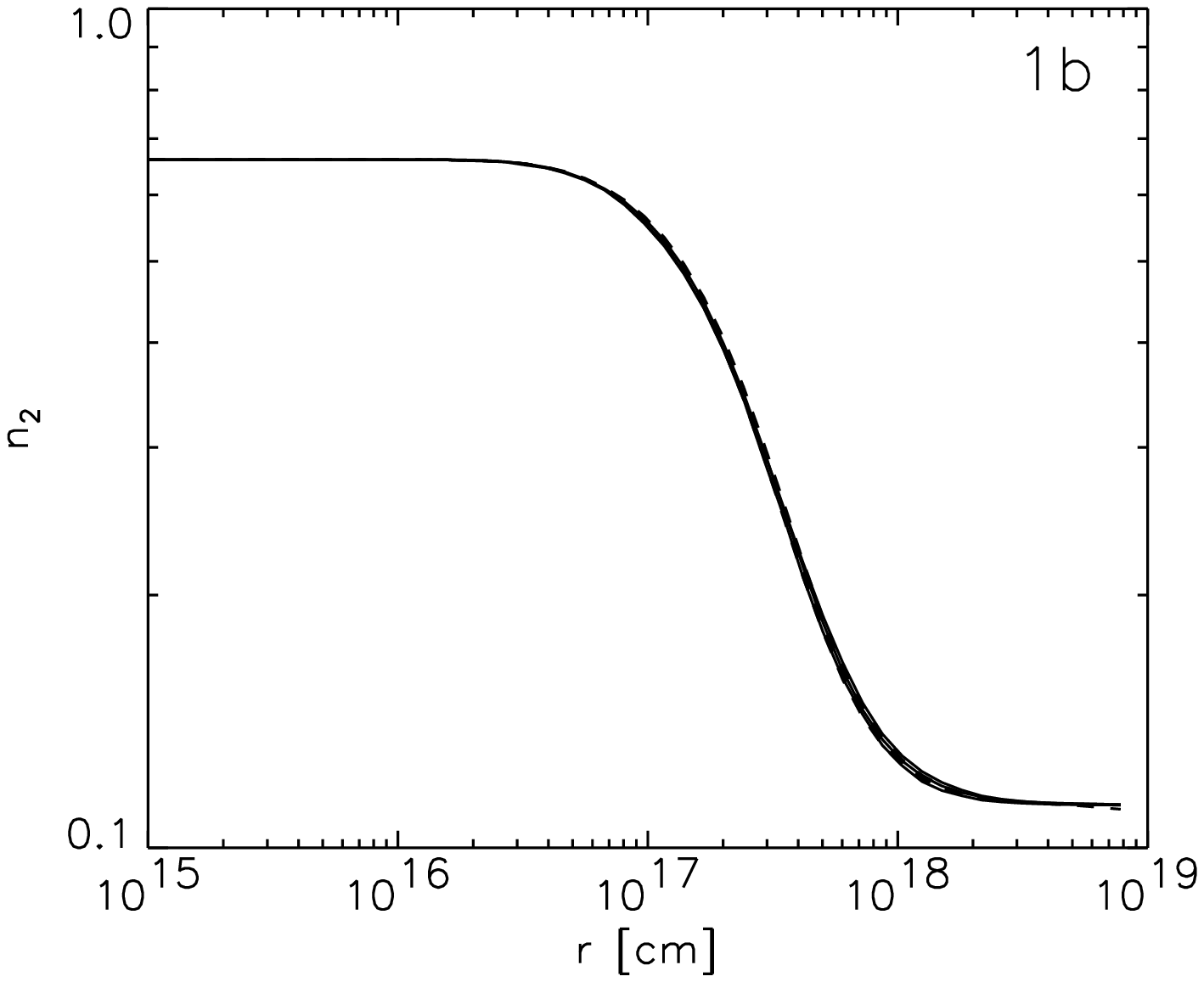}}
\end{center}
\caption{Results of the two-level molecule problems. Shown here are
the populations of the upper level of the 2-level molecule. Top panel:
problem 1a. Bottom panel: problem 1b. Plotted are the (A)MC codes,
denoted by solid lines and the ALI/SMULTI codes using dashed
lines. }\label{fig-tla-n2}
\end{figure}

\begin{figure*}
\resizebox{9cm}{!}{\includegraphics{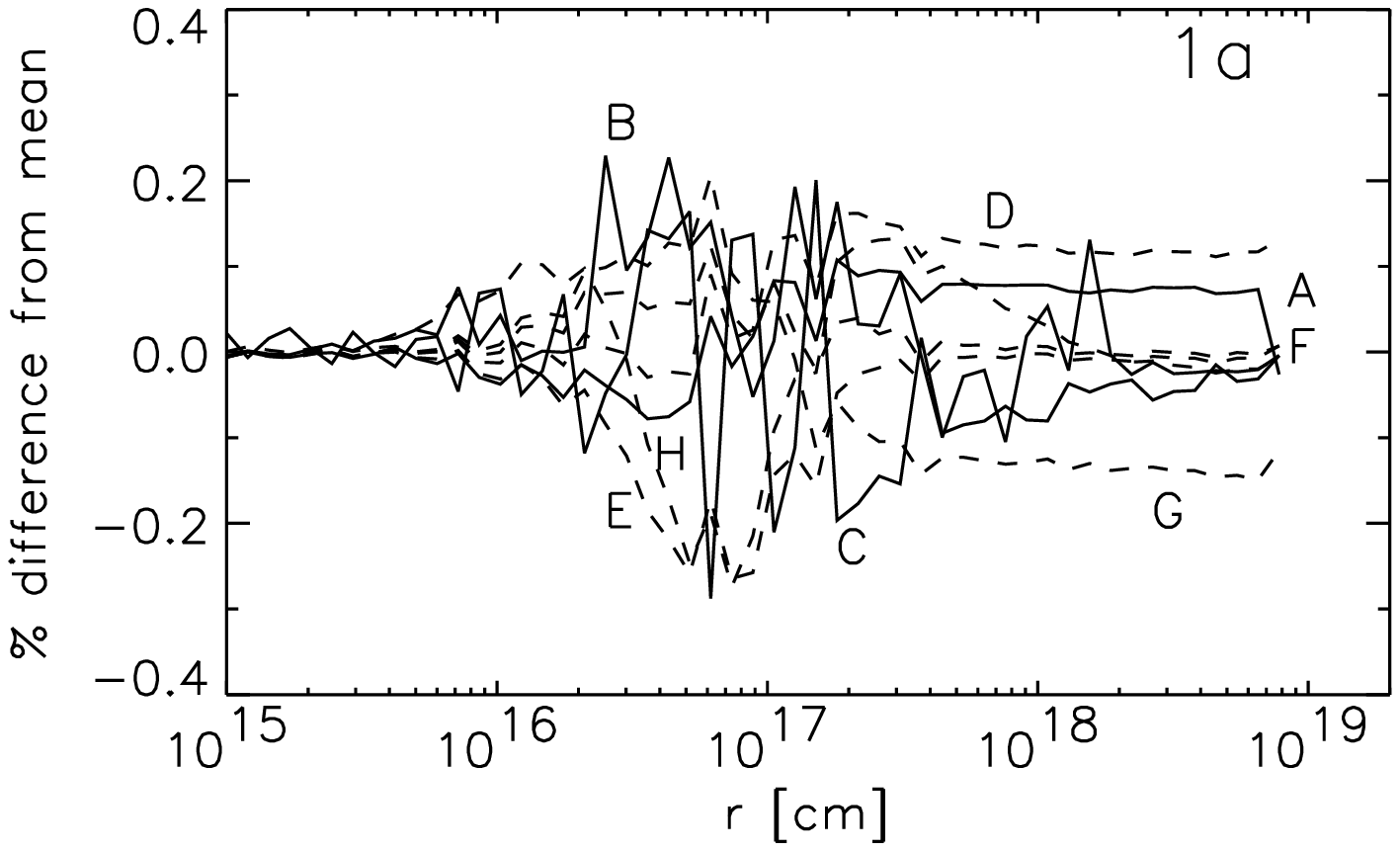}}
\resizebox{9cm}{!}{\includegraphics{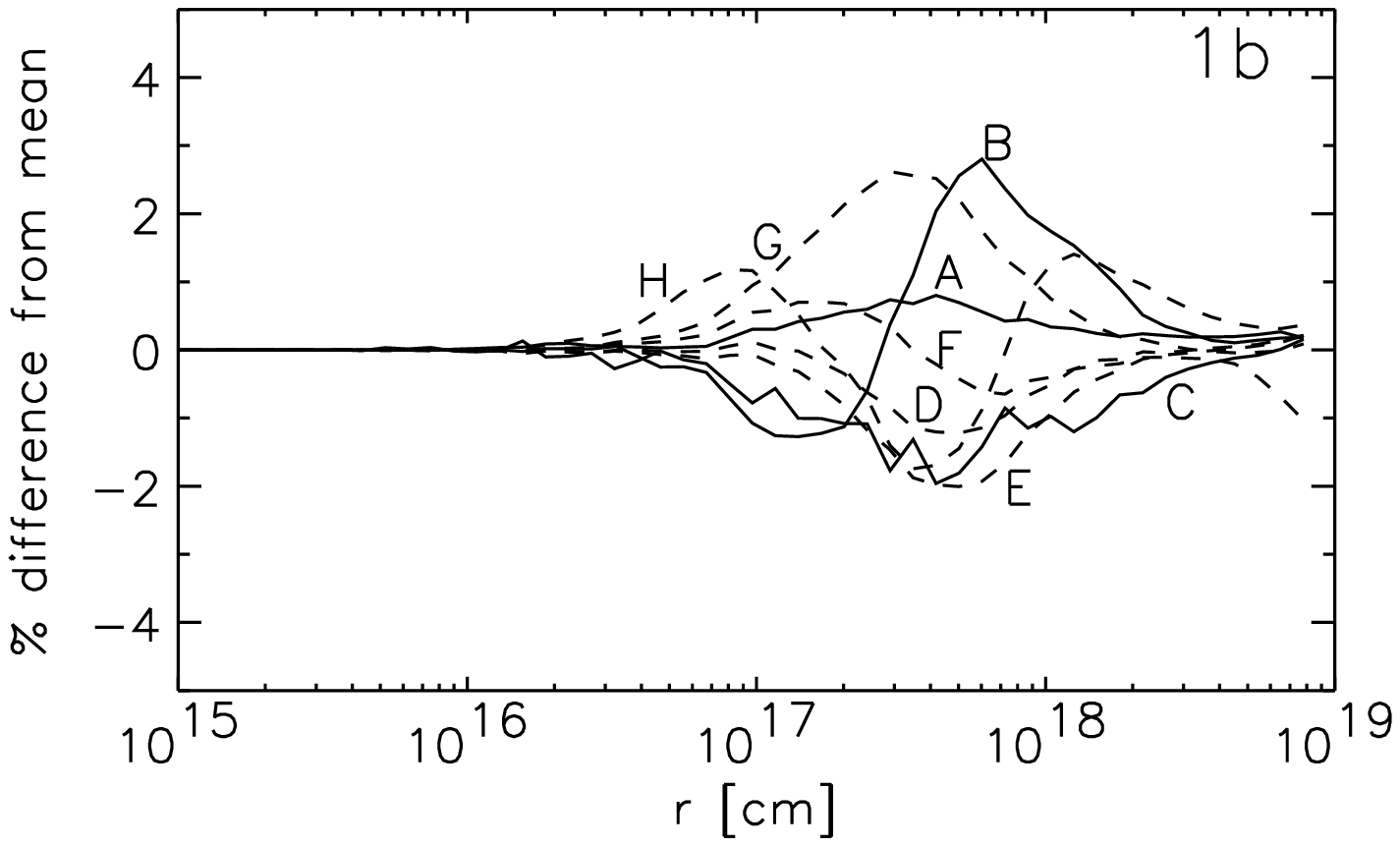}}
\caption{As Fig.\ref{fig-tla-n2}, but now shown as the relative
difference from the average of the results for problems 1a (left) and
1b (right). Plotted are the (A)MC codes, denoted by solid lines and
the ALI/SMULTI codes using dashed lines. The notation for each of the
codes is given in Table 1. Note that the differences between the codes
is of the same order as the Monte Carlo noise for problem 1a,
indicating that the codes have converged to the correct level.}
\label{fig-tla-diffn2}
\end{figure*}

The results for the problems 1a and 1b are shown in
Fig.~\ref{fig-tla-n2}.
The interior of the cloud is completely thermalized ($T_{\rm
ex}=T_{\rm kin}$). The decline of $n_2$ at larger radii is due to the
departure from LTE. At the outer radii the population again saturates
to a constant value, which is due to the microwave background
radiation ($T_{\rm ex}=T_{\rm CMB}$). All codes agree reasonably well,
although small differences remain distinguishable. The relative
differences between the codes are shown in Fig.\ref{fig-tla-diffn2} in
comparison to the mean.

If we consider the 'mean' solution to be the correct one, then the
typical errors are of the order of a few percent or less. The errors
are clearly greater for problem 1b than for problem 1a. This is to be
expected, since problem 1b has higher optical depth and is therefore
numerically more challenging. For problem 1a, the codes show a small
spread with a standard deviation of 0.2 \%. i.e. they agree with each
other to within $2\cdot10^{-3}$ relative difference, comparable to the
accuracy criteria of the codes, clearly visible in the noise for the
Monte Carlo results. For problem 1b the best agreement reached between
the codes shows a standard deviation with a local maximum of 2\%
$(\Delta n_2/n_2\sim 2\cdot10^{-2})$. To achieve maximum accuracy, the
codes used a spatial resolution of $(R_{i+1}-R_{i})/R_i\simeq 0.05$
for this problem, but this high spatial resolution did not reduce the
spread between the codes to the $10^{-3}$ level.

For problem 1b (the most difficult of the two) an ALI code with Ng
acceleration with spatial resolution $(R_{i+1}-R_{i})/R_i\simeq 0.04$
typically takes about 160 iterations to meet the converge criterion of
$\Delta n_2/n_2<10^{-7}$. The same problem, but for a less stringent
convergence criterion ($\Delta n_2/n_2<10^{-4}$) and for a lower
spatial resolution ($(R_{i+1}-R_{i})/R_i\simeq 0.2$) typically takes
about 40 iterations. The relative error introduced by the lower
spatial resolution and lower convergence criterion for this problem is
at most 1\%, provided that second-order (or higher) integration of the
transfer equation is used.  The use of first-order integration
introduces errors of the order of $12\%$, even for the high-resolution
runs. For test problem 1a, at low resolution and with a weak
convergence criterion ($\Delta n_2/n_2<10^{-4}$), an ALI code
typically uses 18 iterations.

For the AMC-type codes with Ng acceleration (Juvela 1997), problem 1b
typically requires about 75 iterations, with about 120 photon packages
(rays) per iteration. It was found that the application of Ng
acceleration at the very start of the iteration had a negative
influence on the convergence rate.  Good convergence was achieved by
applying Ng only after about 25 iterations. As in the case of the ALI
type codes, problem 1a was much more benign: 15 iterations with about
120 photon packages (rays) per iteration.

It should be said that the AMC-type codes allow considerable freedom
in how to disentangle the number of iterations and the number of
photon packages per cell per iteration.  Therefore the numbers given
here may vary from code to code, i.e., in most codes the number of
photon packages per iteration per cell is used. For the accelerated
Monte Carlo codes, acceleration of the convergence only operates when
some of the cells are optically thick. If all cells are optically thin
but the entire source optically thick, the local radiation field
becomes negligible compared to the overall radiation field and
separating local and global contributions no longer effectively speeds
up convergences.

The SMULTI code converged to below 10$^{-4}$ in 30 iterations and to
below 10$^{-5}$ in 50 iterations for the second model.  For the first
model the code converged in 10 iterations.

\subsection{Results of problems 2a/2b}
\subsubsection{Level populations}
\begin{figure}[ht!]
\resizebox{\hsize}{!}{\includegraphics{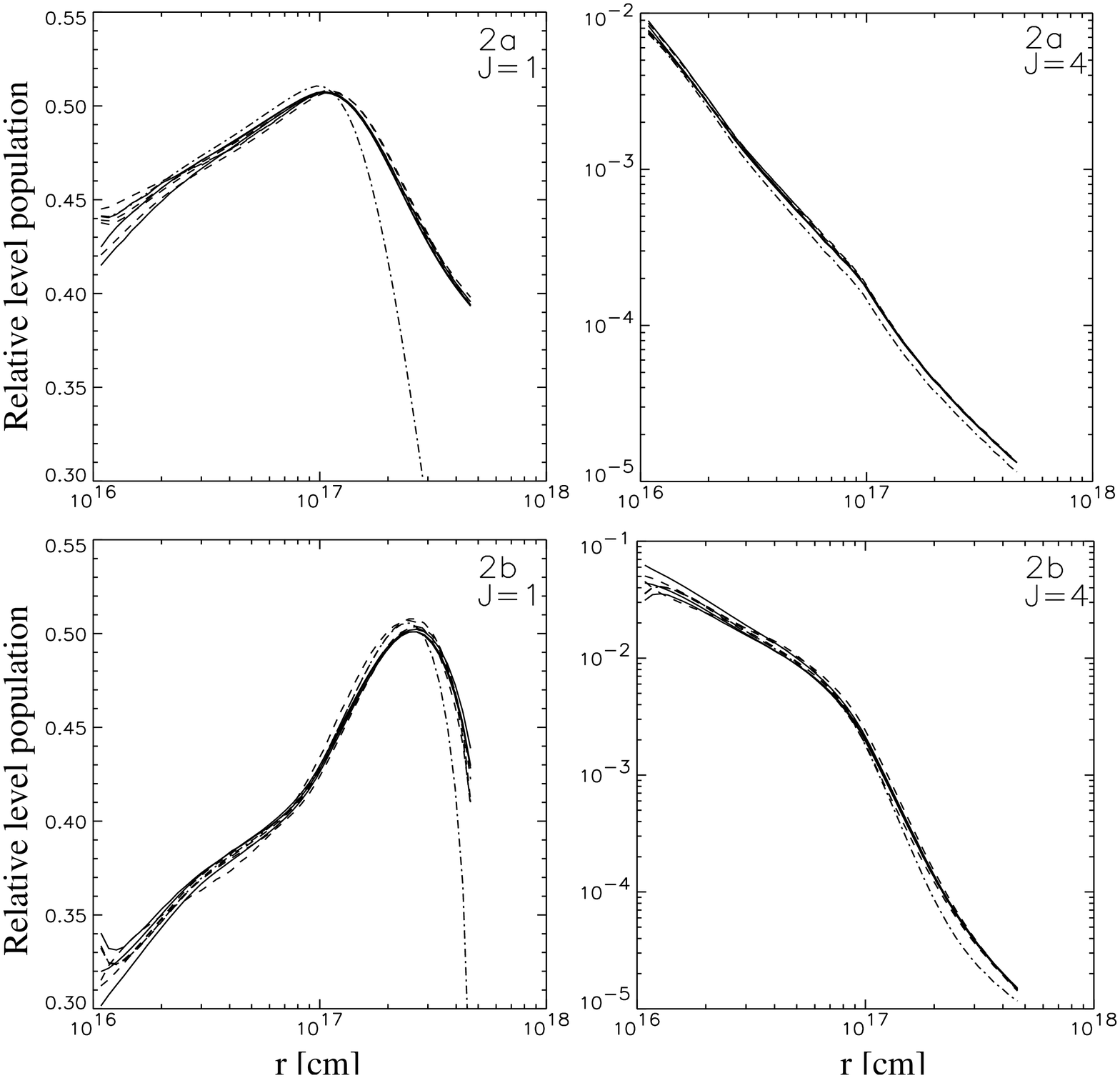}}
\caption{Populations of levels $J$=1 and $J$=4 in the optically thin
(model 2a) and optically thick (model 2b) cases. The dotted line is
the solution where no cosmic background radiation was added, the solid
lines represent the (A)MC-based codes and the dashed lines the
ALI-based codes. Due to the low temperatures and densities of the
problem, most molecules are in the lowest rotational states. Only
codes A--G participated in this comparison.}
\label{fig: pop1}
\end{figure}

The results for the test problems 2a and 2b are shown in
Fig.~\ref{fig: pop1}. The direct comparison is shown for two different
levels of HCO$^{+}$, $J$=1 and $J$=4. These levels were chosen because
they represent the emitting levels of two easily and often observed
lines of the molecule probing different density regimes. The $J$=1--0
line lies at millimeter wavelengths at 89 GHz and the critical density
of the $J$=1 level is $3.4\times 10^4$ cm$^{-3}$.  The $J$=4--3 line
occurs in the sub-millimeter at 356 GHz, with a critical density for
exciting the $J$=4 level of $1.8\times 10^6$ cm$^{-3}$.  The different
model results are plotted on top of each other, together with one
additional calculation where the CMB radiation field was deliberately
ignored. In the level populations, the effect of the CMB is visible as
a lack of excitation of the $J$=1 level in the outer regions. The
figure immediately shows that in the outer region, the excitation at
line center of the $J$=1--0 line is controlled by the CMB field and
not by the local temperature and density. The $J$=1 level lies only
4.3 K above ground, close to the $T_{\mathrm{CMB}}$ (2.728 K)
temperature, so that the 1--0 transition can be effectively excited by
the peak of the CMB radiation.  The density in the outer regions of
the collapsing cloud is too low for the molecule to be excited to LTE.
The $J$=4 level population is less affected by the CMB radiation
field.

\begin{figure*}[ht!]
\resizebox{\hsize}{!}{\includegraphics{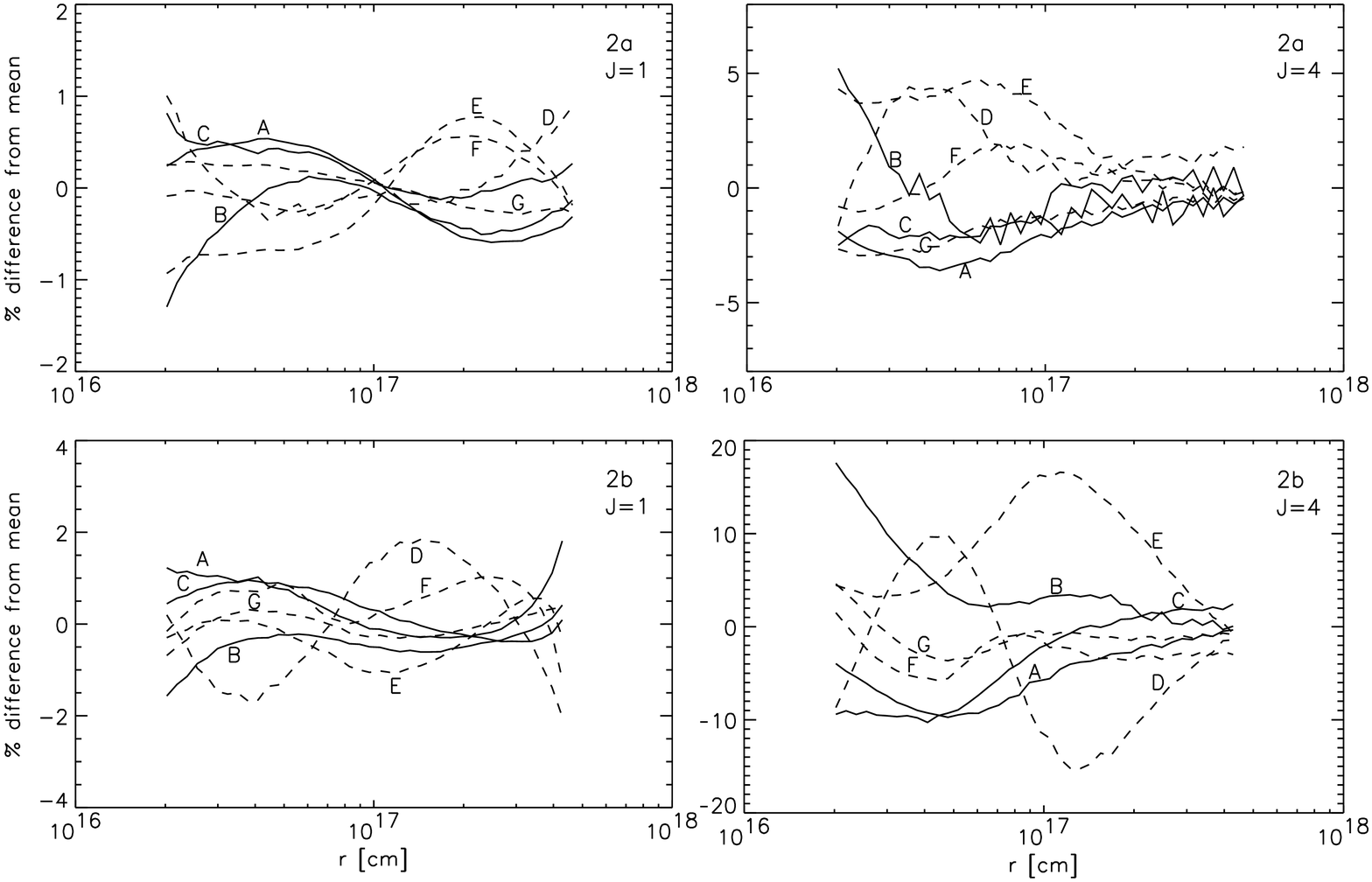}}
\caption{Differences of the populations in the $J$=1 and $J$=4 levels
to the mean for both model 2a and 2b. The solid lines represents the
(A)MC based codes and the dashed lines the ALI based codes.  The
notation of each of the lines is given in Table 1. Only
codes A--G participated in this comparison.}
\label{fig: perc}
\end{figure*}

To show a more quantitative measure of the accuracy of the results,
the level populations are plotted versus their difference to the mean
of all the results (Fig. \ref{fig: perc}). Only results including the
CMB radiation field are used to calculate the mean. At the inner
boundary, the solutions diverge into two main groups. This is a result
from the inner boundary condition adopted by the different authors.
The density and temperature of the test problem were not specified
within the inner radius and were either chosen as an empty sphere or a
non-rotating sphere of constant density. The different solutions near
the inner boundary have a relatively small error (Fig. \ref{fig:
perc}) showing that in this case the inner boundary has little
influence on the overall solution and can be ignored in the
specification of the problem.

\noindent

The calculated populations of the $J$=1 level show a standard
deviation ranging from 2\% close to the star down to 1.5\% for problem
2b and below 1\% in problem 2a. The $J$=4 level has a larger spread
with a standard deviation of 8\% close to the inner boundary up to
12\% at $r$=2$\times10^{17}$ cm and down to 2\% at the outer boundary
for the high $\tau$ (problem 2b) case. The HCO$^{+}$ lines have high
critical densities and the $J$=4 level population is far from LTE. In
addition, the optical depth is high,making the calculation for the
4--3 line particularly difficult. In problem 2a the $J=4$ level rises
from 3\% close to the inner boundary up to 4\% in the transition region
and down to 1\% close to the outer boundary. Lacking an analytical
solution, we take the average of the numerical results as the best
estimate of the exact solution. The deviations from the average
therefore give an estimate of the implicit and explicit approximations
of the codes.  Fitting models to observational data with criteria
better than these deviations will not lead to more accurate estimates
of the physical parameters. 

For all the problems, the level populations were also calculated using LI
by Dullemond and Wiesemeyer. The populations agreed within the errors as
shown in Figs. 4 \& 6 for the low $\tau$ cases. However in both the 1b and
2b problems the solution was not reached using similar convergence criteria
as the ALI calculations. Even though the LI method itself should in
principle find the same solution, the number of iterations and more strict
convergence criteria make it a numerically too costly task to solve.

The problem is particularly sensitive to the gridding of the physical
parameters. A comparison of two runs, where one had twice as many
gridcells, showed a significant decrease of the error compared to the
mean value for one of the Monte Carlo code (C). This shows the
importance of a correct gridding of the problem.

\noindent

\begin{figure}[ht!]
\resizebox{\hsize}{!}{\includegraphics{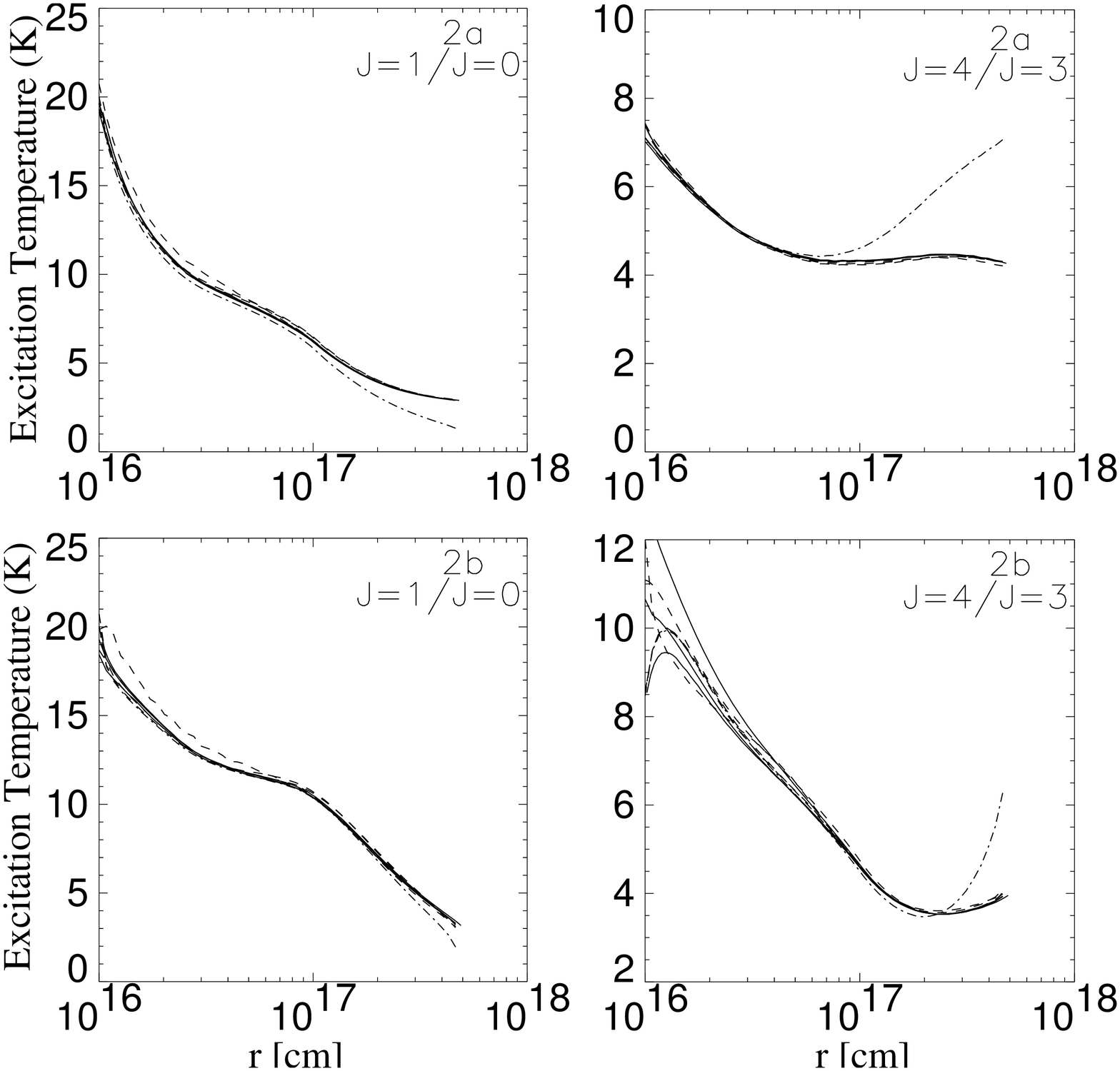}}
\caption{Excitation temperatures [K], as defined in Eq.~(\ref{eq:
tex}), for the levels $J$=1/$J$=0 and $J$=4/$J$=3.  Top panels: low
optical depth case (problem 2a); bottom panels: high optical depth
case (problem 2b). The dotted line indicates the model result with the
CMB radiation neglected, the solid lines represent the (A)MC-based
codes and the dashed lines the ALI-based codes. When the lines are in
LTE, the excitation temperature is equal to the kinetic temperature
(Figure \ref{fig: model}).}
\label{fig: tex}
\end{figure}
\subsubsection{Excitation temperature}
As a second comparison, the excitation temperature (Fig.~\ref{fig:
tex}) is plotted for all solutions, which may be a more intuitive
manner to show the results. When level populations are in LTE, the excitation
temperature equals the local kinetic temperature. Comparing the results
with Fig. \ref{fig: model}b, the excitation temperature
of the 1--0 line is
close to LTE near the inner boundary but becomes
subthermal at larger radii. The $J$=4--3 excitation temperature is always
well below its LTE value, which is due to its much higher critical density.
As Fig. \ref{fig: tex} shows, the
excitation temperature never drops below 2.728 K, except in the case
where the CMB radiation was ignored. In the inner region ($r< 5\times10^{16}$
cm), the excitation of the $J$=1 level is dominated by collisions, while
in the outer parts, it is dominated by radiation. The transition region
between these two extremes shows the largest differences in the
results (Fig. \ref{fig: perc}a).

\subsubsection{Line profiles}
\begin{figure}
\resizebox{\hsize}{!}{\includegraphics{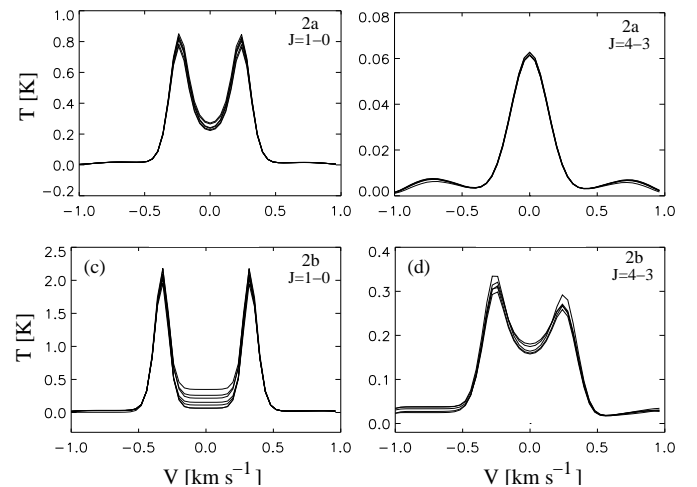}}
\caption{Calculated intensity [K] of the $J$=1--0 and $J$=4--3 lines
 for the problems 2a and 2b. In the high optical depth case (lower two
 panels), the blue-shift or `infall asymmetry' is visible. The low
 resolution of the spatial grid becomes apparent in the line emission
 at $\pm$ 0.75 km s$^{-1}$.  }
\label{fig: lines}
\end{figure}

Figure \ref{fig: lines} compares the results in terms of the line
profiles. The calculated level populations at each position in the
cloud are used to compute the velocity profiles of the selected lines
using a program which calculates the sky brightness distribution. To
ensure an equivalent emitting mass, the inner sphere is assumed
to be empty for all models. The profiles are calculated by
constructing a plane through the origin of the cloud perpendicular to
the line of sight, with a spatial resolution small enough to sample
the physical distributions.  A ray-tracing calculation is performed
through this plane from $-\infty$ to $+\infty$, keeping track of the
intensity in the different velocity bins. For this calculation the
program SKY was used, part of the RATRAN code {\tt
(http://talisker.as.arizona.edu/$\sim$michiel/)} By applying a single
common code for these calculations, no secondary uncertainties are
introduced. The resulting sky brightness distribution is convolved
with a beam of 14$\arcsec$ for the $J$=4--3 transition, appropriate
for the James Clerk Maxwell Telescope (JCMT) at this frequency.  For
the $J$=1--0 transition, the beam size is taken to be 29$\arcsec$
representative of the IRAM 30m telescope. The distance is assumed to
be 250 pc.

Comparison of Fig.~\ref{fig: lines}a and c shows that increasing the
abundance of HCO$^{+}$ by a factor of 10 changes the $J$=1--0 peak
intensity by only a factor of 2.5. The $J$=1--0 line becomes more
self-absorbed, indicating that the line has indeed become more
optically thick. The $J$=4--3 emission line is optically thin for the
low abundance, as seen from its single-peaked profile in
Fig.~\ref{fig: lines}b. The more optically thick model shows the
characteristic asymmetric line structure, skewed to the blue. The
differences in the level populations are visible in the line profiles.
In the $J$=1--0 line profiles (Fig. \ref{fig: lines}c) all solutions
lie on top of each other in the line wings and only a slight
difference can be seen at line center. The error is largest in the
region with low velocities represented by the larger errors in the
center of the line profile.  The integrated intensity profiles differ
by a few \% for low $\tau$ and 6-7\% for the high $\tau$ case.  The
error is in general larger for the $J$=4 level, which shows up as a
larger error in the integrated lines for the $J$=4--3 transition.  For
low $\tau$ the error is $\approx$ 2\%, but for high optical depths
deviations up to 12\% are found for the most outlying case. Although
substantial, these differences are generally less than the current
calibration uncertainties of the observational data, which are
typically 20\% to 30\%.

 The differences in the profiles shown here give a rough indication of
the quality of fitting necessary to interpret observational data. It
should be noted that some differences are likely due to different
implementations of the cloud model.  One can fit a model precisely to
an observational data set using a particular radiative transfer code;
however the results will not necessarily be the same if a different
code is used. The derivation of physical parameters is therefore
always limited in accuracy to the error bars given by the code itself
in comparison to other codes.
\noindent

\section{Discussion and conclusion} 
In spite of the fact that the spread in our solutions lies within
the accuracy limits of current-day (sub-)millimeter instruments, it is
important to understand how this spread comes about. The first two
test cases were defined analytically, and avoided difficulties of
gridding as much as possible. The small spread of these results, even
at high optical depths, seems to show that in principle the radiative
transfer is done correctly by all codes. The larger spread in the
HCO$^{+}$ collapsing cloud problem can therefore be attributed to the
problem of gridding, both in space and in frequency. The coarse
spatial gridding in the setup of the problem, and the presence of
strong velocity gradients makes it very likely that different
interpretations of the sub-grid behavior of temperature, density and
velocity lead to slightly different results.

The main results of the comparison of radiative transfer codes for
molecular lines can be summarized as follows.

\begin{itemize}

\item{All relevant methods currently used in molecular astrophysics
agree to a few \% for optical depths up to $\tau\sim$4800. At high
optical depth and in transition layers between collision-dominated and
radiation-dominated excitation, relative differences up to 2 \% can
arise for well defined ``simple'' problems. In practice, for more
complex models, the relative differences are higher due to gridding
and geometry problems, reaching relative differences up to 12 \%
locally for the high optical depth model described in this paper
(problem 2b) . However, in most practical applications in molecular
astrophysics the optical depths are lower than in our test problems,
and therefore this 12\% error can be regarded as an upper limit.  }

\item{The error bars on the line profiles are generally much less than
10\%, and therefore lie within the calibration errors of typical
(sub-)mm observations. Only in the worst case these errors may be
comparable to the observational uncertainty.}

\item{The choice of gridding is of extreme importance, and is one of
the major causes of the deviations in the problems presented. Special
care should be taken when constructing models for specific problems.}

\item{Abundances and excitation temperatures derived from lines which
are formed in the transition layer should be interpreted with 
caution.}

\item{The direct comparison of results of different programs speeds up
significantly the debugging process of new programs.}

\end{itemize}

The work presented here is one step in the direction of
standardization of radiative transfer computations in molecular
rotational lines. Suggestions for future work in this area include:
(1) establishment of a database for collisional rate coefficients,
accessible to the community through the WWW; (2) a campaign to have
inelastic rate coefficients measured or calculated for those molecules
for which these data are unavailable, and (3) comparison of 2D
radiative transfer codes, with a test problem based on, e.g., the
rotating flattened collapse described by Terebey et al.\ (1984) or the
``sheet'' models of protostellar collapse by Hartmann et al.\
(1996). These developments will be vital to interpreting the
high-quality data which e.g.~HIFI (the heterodyne instrument onboard
the Herschel Space Observatory) and ALMA will provide.

\begin{acknowledgements}
We are grateful to the anonymous referee
for suggestions and comments that improved the paper considerably.
We thank Ewine van Dishoeck for useful advice and discussions during
this work.  The comparison of line radiative transfer codes was
started during a workshop in Leiden in May 1999 in the Lorentz
Center. Financial support for the workshop from the Lorentz Center,
the Netherlands Research School for Astronomy (NOVA), the Leidsch
Kerkhoven Bosscha Fund (LKBF), the Netherlands Organization for
Scientific Research (NWO) and the UK Particle Physics and Astronomy
Research Council (PPARC) is gratefully acknowledged.  SDD acknowledges
support from the Research Corporation.
\end{acknowledgements}

\end{document}